\documentclass[10pt,twocolumn,letterpaper]{article}
\usepackage{mdwmath}
\usepackage{mdwtab}
\usepackage{eqparbox}
\usepackage{amsmath,amssymb,amsfonts}
\usepackage{caption}
\usepackage{algpseudocode}
\usepackage{algorithm}
 \usepackage{cvpr}              % To produce the CAMERA-READY version
\usepackage[dvipsnames]{xcolor}

\definecolor{cvprblue}{rgb}{0.21,0.49,0.74}
\usepackage[pagebackref,breaklinks,colorlinks,citecolor=cvprblue]{hyperref}
\usepackage[none]{hyphenat}
\def\paperID{11976} % *** Enter the Paper ID here
\def\confName{CVPR}
\def\confYear{2024}

\title{DAP: A Dynamic Adversarial Patch for Evading Person Detectors}

\author{%
 Amira Guesmi$^{1}$, Ruitian Ding$^{2}$, Muhammad Abdullah Hanif$^1$, \\ 
Ihsen Alouani$^3$, Muhammad Shafique$^1$ \\
$^1$ eBrain Lab, New York University (NYU) Abu Dhabi, UAE \\ $^2$ NYU Tandon School of Engineering, USA\\ %Division of Engineering, 
$^3$ CSIT, Queen’s University Belfast, UK\\
}

\begin{document}
\maketitle
%%%%%%%%% ABSTRACT
\begin{abstract}
Patch-based adversarial attacks were proven to compromise the robustness and reliability of computer vision systems.
However, their conspicuous and easily detectable nature challenge their practicality in real-world setting. To address this, recent work has proposed using Generative Adversarial Networks (GANs) to generate naturalistic patches that may not attract human attention. However, such approaches suffer from a limited latent space making it challenging to produce a patch that is efficient, stealthy, and robust to multiple real-world transformations.
This paper introduces a novel approach that produces a Dynamic Adversarial Patch (DAP) designed to overcome these limitations. DAP maintains a naturalistic appearance while optimizing attack efficiency and robustness to real-world transformations.
The approach involves redefining the optimization problem and introducing a novel objective function that incorporates a similarity metric to guide the patch's creation. Unlike GAN-based techniques, the DAP directly modifies pixel values within the patch, providing increased flexibility and adaptability to multiple transformations. Furthermore, most clothing-based physical attacks assume static objects and ignore the possible transformations caused by non-rigid deformation due to changes in a person’s pose. To address this limitation, a `Creases Transformation' (CT) block is introduced, enhancing the patch's resilience to a variety of real-world distortions.
Experimental results demonstrate that the proposed approach outperforms state-of-the-art attacks, achieving a success rate of up to 82.28\% in the digital world when targeting the YOLOv7 detector and 65\% in the physical world when targeting YOLOv3tiny detector deployed in edge-based smart cameras. %This work advances the field of adversarial patch generation, offering an inconspicuous yet highly effective solution for real-world adversarial attacks.
\end{abstract}    
\section{Introduction}
\label{sec:intro}
\begin{figure}[!ht]
\centering %, height=2.5cm
\includegraphics[width=\columnwidth]{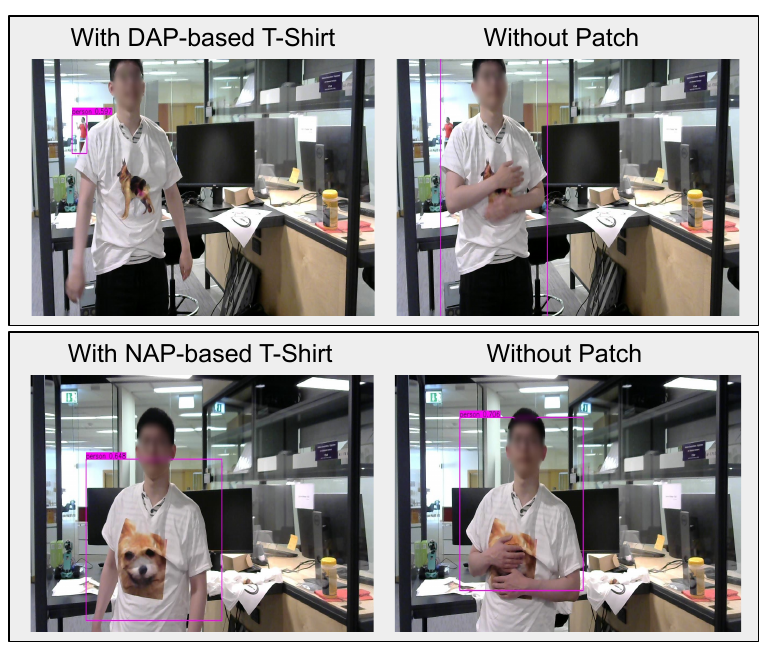}
\caption{Illustration of different Adversarial T-shirts against Yolo detector. DAP-based t-shirt (ours) is still effective in the presence of non-rigid deformations compared to the GAN-based t-shirt (NAP) \cite{Hu21}. }
\label{DAPvsNAP}
\end{figure}
%In recent years, deep learning (DL) models have been increasingly used to tackle real-life problems, thanks to their ability to learn features automatically through multiple levels of abstraction. 
Deep Neural Networks (DNNs) have demonstrated remarkable performance for various real-world applications and are now commonly used in safety-critical and security-sensitive domains such as video surveillance \cite{boudjit2022human,sabater2020robust,8524260}, self-driving cars \cite{al2017deep}, and healthcare \cite{miotto2018deep}. 

\begin{figure*}[!htp]
\centering %, height=2.5cm
\includegraphics[width=0.9\textwidth]{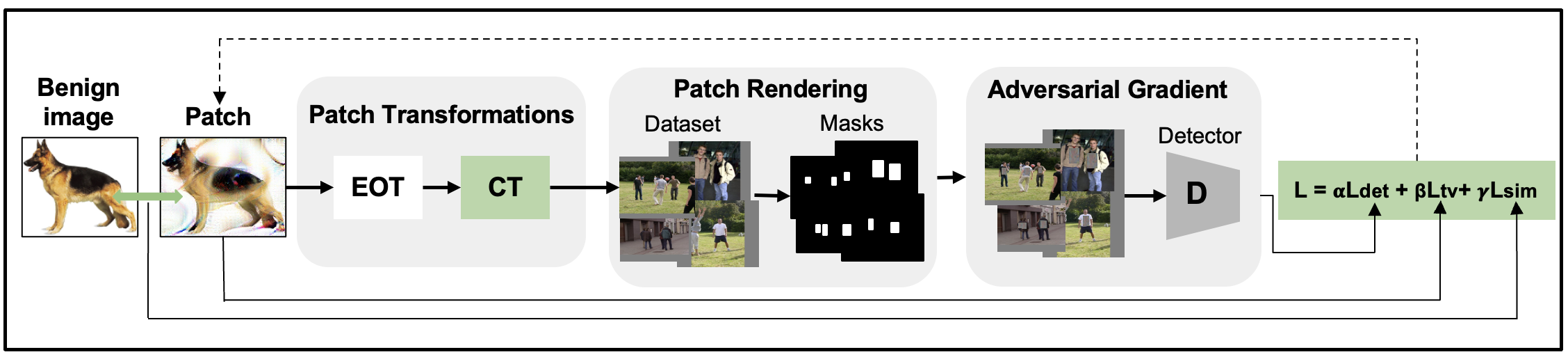}
\caption{Overview of our dynamic adversarial patch generation framework which crafts patches that can be printed on a T-shirt and evade object detectors under different real-world conditions. }
\label{approach}
\end{figure*}
However, studies have shown that DNNs are vulnerable to adversarial perturbations \cite{CW, LiV15, adv_example, PapernotMJFCS15}. These malicious examples can even be realized physically and deployed in the real world, posing a serious security/safety threats.

There are two types of attack settings: digital attacks and physical attacks. In digital attacks \cite{fgsm, CW}, an attacker adds adversarial noise to the digital input image, optimized to be undetectable by human eyes while monitoring the noise budget constraint in the generation process. In contrast, in physical attacks \cite{phy9,guesmi2023advrain, phy10}, the attacker designs adversarial patches that are printable in the physical world and deploys them in the scene captured by the victim model. These patches are not generated under noise magnitude constraints but rather are generated under location and printability constraints, making physical patches-based attacks more practical and damaging for real-life scenarios.
%\begin{figure}
%\includegraphics[width=\columnwidth]{figures/threat.png} %, height=2.5cm
%\caption{Threat model of the proposed attack.}\label{threat}
%\end{figure}
However, current adversarial patches are limited in the following ways:

\begin{figure}
\includegraphics[width=0.5\textwidth]{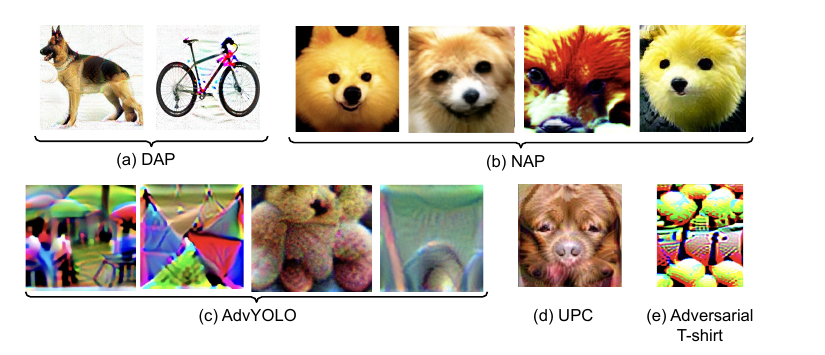} %, height=2.5cm
\caption{DAP vs State-of-the-Art patches: (a) DAP, (b) NAP (for YOLOv3tiny) \cite{Hu21}, (c) Adversarial patch \cite{thys2019}, (d) UPC \cite{Huang2020}, and (e) Adversarial T-shirt \cite{adversarialtshirt}. Our patch is more natural looking and less conspicuous than others so it is harder to human observers to identify it.}\label{DAP}
\end{figure}

\textbf{Conspicuous patterns.} Previous research on adversarial patches for object/person detection \cite{Eykholt2018, thys2019, Zhao2019} has largely focused on improving attack performance by increasing the strength of the adversarial noise. However, this approach often results in patches that are easily identifiable by human observers, limiting their effectiveness in real-world scenarios \cite{Pavlitskaya22,Bai22,Hu21,Liu20}. To address this issue, some researchers have proposed using generative adversarial networks (GANs) to generate more natural-looking patches. While promising, these approaches can be inefficient (i.e., result in low attack success rate as depicted in Figure \ref{DAPvsNAP}) and may not always converge to realistic/natural-looking or efficient patterns (refer to Section \ref{sec:limitations} for further details).

\textbf{Assumption of static objects.} Previous research assumed static objects \cite{thys2019,Huang2020, Hu21}, however, when printing an adversarial patch on a T-shirt, it is crucial to ensure that the patch is robust to additional transformations caused by changes in the fabric (e.g., deformation and orientation changes) due to a person's movements. In particular, the constantly changing wrinkles and creases in the fabric can significantly affect the effectiveness of the attack. We demonstrate that the GAN-based techniques suffer from a limited latent vector which makes incorporating multiple transformations at once very challenging (See Figure \ref{DAPvsNAP}).

In this paper, we propose a novel framework (See Figure \ref{approach}) to generate natural-looking adversarial patches for performing effective and robust adversarial attacks on object/person detection systems, by guiding the optimization process towards a target natural image and simultaneously maximizing the victim model's loss. To achieve this, we propose a novel objective function that includes a similarity loss to guide the pattern of the generated noise, resulting in natural-looking patches  (see Figure \ref{DAP}). Additionally, we incorporate a crease transformation block to model possible deformations that could occur to the adversarial patch when used to conceal dynamic objects (See Figure \ref{DAPvsNAP}).

\noindent\textbf{Contributions -- } The main contributions of this paper are summarized as follows:
\begin{itemize}
    %\item To the best of our knowledge, we first to propose a patch that solves the trilemma of efficiency, stealthiness, and robustness. 
    \item We investigate the limitations of GAN-based approaches when used to generate robust naturalistic adversarial patterns, and we show that these techniques can not incorporate multiple transformations while maintaining high performance due to the limited latent space compared to the high flexibility and the larger space provided by our approach as it relies on directly manipulating the patch pixels.
    
    %\textcolor{blue}{Is there any new knowledge out of this investigation? If yes, it is worth to mention the outcome} %(Section \ref{limitations}).
    \item We propose a framework (See Figure \ref{approach}) that generates GAN-free naturalistic patches that can resemble any predefined pattern, while maintaining high attack success rate under multiple transformations (e.g., clothing creases and wrinkles, random noise, brightness and contrast variations, re-scaling, and rotation, etc.). 
    \item To increase robustness against non-rigid deformations experienced by a printed adversarial patch on a T-shirt and caused by pose changes of a moving person, we develop a Creases Transformation (CT) block that models these non-rigid deformations by compressing the pixels following a randomly selected direction.
    %We propose a framework (See Figure \ref{approach}) that generates \textcolor{blue}{GAN-free} naturalistic patches that can resemble any predefined pattern, while maintaining high attack success rate under multiple transformations (e.g., random noise, brightness and contrast variations, re-scaling, rotation, and creases, etc.). %(Section \ref{sec:proposed}).
    %\item We propose leveraging similarity metrics to force the optimized noise to follow a predefined pattern.
    %\item We develop a Creases Transformation (CT) block to model the non-rigid deformations experienced by a printed adversarial patch on a T-shirt and caused by pose changes of a moving person. We also show the importance of such non-rigid transformation to ensure the effectiveness of adversarial T-shirts in the physical world. 
    
    %\textcolor{blue}{bare in mind , this community is more interested in theory/knowledge than in practical attacks like security venues -- try to rephrase things accordingly}
    %\item We propose a patch that can be printed on a T-shirt and still be robust to different real-world transformations and wrinkles (Section \ref{wrinkles}). 
    \item We thoroughly examine the performance/attack success rate of the proposed method in terms of mean average precision (mAP) both with and without transformations, and transferability between detectors.  Our patch achieves an attack success rate of 82.28\% in the digital world (INRIA dataset) when targeting the YOLOv7 detector and 65\% in the physical world when attacking the YOLOv3tiny detector for edge systems. %We demonstrate the effectiveness of our patch in the digital world, when printed on a paper, and when printed on a T-shirt.  
    %\item We provide two demo videos in the supplementary material to showcase the effectiveness of our dynamic adversarial Patch (DAP)-based T-shirt and that of the GAN-based T-shirt in attacking the Yolo object detectors. 
    %\textcolor{blue}{Merge what is mergeable; last 2 points for example} % (Sections \ref{sec:eval} \& \ref{physical}).
    %add transferability results
    %\footnote{are available at: https://drive.google.com/drive/folders/14Ij\_EBJcrHV\\3bB3\_qGBbUH3sPuQuBAm0?usp=sharing}
    %add real-world experiment results
\end{itemize}

\section{Limitations of GAN-based techniques} %Motivation: 
\label{sec:limitations}

%\textcolor{red}{Summarize this section}

\begin{figure*}[!ht]
\includegraphics[width=2\columnwidth]{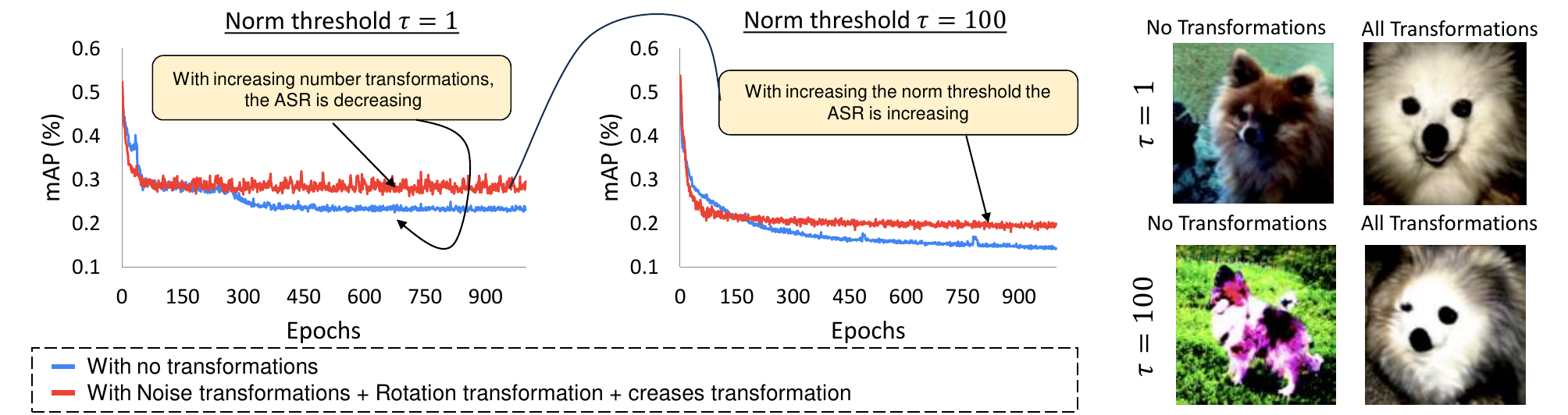}
\caption{\label{Gan_performance_tiny}Mean Average Precision (mAP) convergence curves when training a GAN-based technique with and without different transformations. Illustrating the impact of adjusting the latent vector constraints on attack success rate (ASR) (\textit{Left:} $\tau = 1$ and \textit{Right:} $\tau = 100$) and the corresponding generated patch.}
\end{figure*} 

%Authors in \cite{Hu21, doan2022tnt, Pavlitskaya22} proposed solving the problem of searching for naturalistic patches with adversarial effects by indirectly manipulating the latent vector $z$ of a Generative Adversarial Network that has learnt to approach the natural patch distribution. Although GANs are known to be effective in generating images from random noise, they may suffer from multiple limitations when used to produce naturalistic patches aimed to fool object detectors that can withstand real-world distortions. In this section, we investigate those limitations.
The generator in a GAN is typically trained by sampling random vectors from a standard normal distribution, which results in a high density region centered around the origin. Thus, if the latent vector $z$ is closer to the origin, there is a higher probability of generating realistic images, that is why a constraint $\tau$ on the norm of the latent vector $z$ is required. 
In this section, we investigate the impact of incorporating multiple transformations in the naturalistic adversarial patch generation process and the impact of the norm threshold of the latent vector $z$ on the effectiveness of GAN-based techniques for generating adversarial patches that are robust against real-world transformations.

%To evaluate the GAN-based techniques in generating naturalistic patches, we target the Yolov3tiny detector and we conducted a series of experiments. Our primary focus was on comparing the effectiveness of attacks when no transformations were applied versus a combination of all transformations (including Noise, Rotation, and Creases). Additionally, we explored various norm thresholds. By systematically varying the combinations of transformations applied to the patches, we could gauge their impact on the patch's effectiveness in deceiving the Yolov3tiny detector. This consistent trend in our findings is further supported by additional results presented in the supplementary material for the Yolov4tiny detector.
To assess GAN-based techniques for generating naturalistic patches, we conducted experiments targeting the Yolov3tiny detector. Our emphasis was on comparing attack effectiveness under different conditions: no transformations vs. all transformations (including Noise, Rotation, and Creases). We also investigated various norm thresholds. This systematic exploration allowed us to assess how different transformation combinations influenced patch effectiveness in deceiving the Yolov3tiny detector, with consistent trends also noted in supplementary material for the Yolov4tiny detector. Specifically, as we incorporated more transformations into the generation process, we noticed a decrease in the effectiveness of the generated patch (See Figure \ref{Gan_performance_tiny} and Table \ref{Gan_performance_out}). This observation suggests that there exists a trade-off between the number of transformations applied to the patch and its ability to deceive the detector effectively. While transformations can enhance the patch's camouflage and resilience against real-world transformations, an excessive number of transformations may introduce distortions that hinder its effectiveness.

%Additionally, we explored the influence of adjusting the norm threshold of the latent vector $z$ on the generated patch performance and appearance. By fine-tuning the norm thresholds, we observed an improvement in the effectiveness of the generated patches. Higher norm thresholds, i.e., $\tau = 100$, allowed for larger space, increasing the patch's ability to evade detection. However, it is important to note that as we increased the norm thresholds, the generated patches began to exhibit unrealistic and visually conspicuous features (See Figure \ref{Gan_performance_tiny}). These unrealistic attributes could potentially compromise the patch's stealthiness and raise suspicions, limiting its practicality in real-world scenarios.

%With higher threshold, the optimization process can produce latent vectors $z$ that are not contained within the high density region learned by the generator. Consequently, we cannot expect the generated images to look realistic. However, with lower threshold, the generated patch may look naturalistic but less effective to fool the victim detector and results in a limited latent space which decreases the tolerance to incorporating multiple transformations.

Furthermore, we explored the impact of adjusting the norm threshold for the latent vector $z$ on the performance and appearance of the generated patches. Through fine-tuning these thresholds, we observed enhancements in the effectiveness of the generated patches. Higher norm thresholds, such as $\tau = 100$, allowed for larger space, increasing the patch's effectiveness as shown in Table \ref{Gan_performance_out}. However, it's essential to highlight that as we pushed the norm thresholds higher, the generated patches began to manifest unrealistic and visually conspicuous characteristics (as depicted in Figure \ref{Gan_performance_tiny}). These unrealistic attributes could potentially compromise the patch's ability to remain inconspicuous and might raise suspicions, limiting its practicality in real-world scenarios.
\begin{table}[htp]
\centering
\small
  \begin{tabular}{lcc}
    \toprule
    \textbf{Transformations} & \textbf{$\tau = 1$}   & \textbf{$\tau = 100$}   \\
    \midrule
      \textbf{No transformation}            &  23.43$\%$  &  13.21$\%$  \\
      \textbf{Noise + Rotation + Creases}   & 29.82$\%$    & 21.53$\%$ \\
  \bottomrule
\end{tabular}
\captionof{table}{\label{Gan_performance_out} mAP of GAN-based technique when training using different transformations.}
\end{table}
However, with lower thresholds, the generated patches appear more naturalistic but are less effective at deceiving the target detector. This also results in a narrower latent space, reducing the flexibility to incorporate multiple transformations.
%These experiments yielded valuable insights into the limitations of the GAN-based technique employed to generate naturalistic adversarial patches. Through rigorous testing and evaluation, we were able to identify certain constraints and drawbacks associated with the GAN-based approach. These findings shed light on the challenges and potential shortcomings of using GANs for generating adversarial patches.
%It is worth noting that the tested GANs were not specifically trained on images augmented with the creases deformations and all the other transformations. We believe that the use of such data augmentation techniques for GAN training could potentially introduce further challenges and distortions in the generated samples, as suggested in the literature \cite{tran2021data}, which could result in even less naturalistic patches.
It is worth noting that the tested GANs were not specifically trained on images augmented with these transformations, such as creases deformations. It was shown that training GANs with multiple data augmentation techniques could introduce additional challenges and distortions in the generated samples, potentially making them less naturalistic \cite{tran2021data}.
Further limitations of the existing GAN-based approaches, such as failure to converge, are discussed in the supplementary material.

%\begin{figure}[htp]
%\centering
%\includegraphics[width=0.9\columnwidth]{figures/limited_z_patches.pdf}
%\captionof{figure}{\label{Gan_patch}Generated patches for different latent vector norm threshold.}
%\end{figure}

%\begin{wrapfigure}{l}{9cm}
%  \centering
%\small
%  \begin{tabular}{lccc}
%    \toprule
%    \textbf{Transformations} & \textbf{$\tau = 1$}  & \textbf{$\tau = 50$} & \textbf{$\tau = 100$}   \\
%    \midrule
%      \textbf{No transformation}            &  52.81$\%$ & 47.57$\%$  &  49.67$\%$  \\
%      \textbf{Noise}                        &  52.55$\%$ & 48.51$\%$  &  49.60$\%$ \\
%      \textbf{Noise + Rotation}             & 54.05$\%$  & 50.44$\%$  & 49.99$\%$  \\
%      \textbf{Noise + Rotation + Creases}   & 55.16$\%$  & 52.07$\%$  & 51.76$\%$ \\
%  \bottomrule
%\end{tabular}
%   \caption{\label{Gan_performance} mAP of GAN-based technique when training using different transformations.}
%\end{table}
%\end{wrapfigure}
%\begin{wrapfigure}{l}{6cm}
%\includegraphics[width=6cm, height=2.5cm]{figures/limited_z_patches.pdf}
%\caption{Generated patches for different latent vector norm threshold.}\label{Gan_patch}
%\end{wrapfigure}

\section{Proposed Approach}
\label{sec:proposed}
In this paper, we propose an attack that simultaneously satisfies the three key requirements needed for the adoption of adversarial attacks in the real world. These requirements include: \textit{\textbf{Effectiveness}} in degrading the person detector’s performance. \textit{\textbf{Stealthiness}} against human visual inspection (i.e., being unrecognizable by the observer). \textit{\textbf{Robustness}} in maintaining attack ability in a dynamic environment including robustness to physical constraints.
\subsection{Attack Effectiveness}
%============================
\label{Ldet}
Figure \ref{approach} represents an overview of the proposed framework.
Our goal is to generate physical adversarial patches that are natural-looking while still maintaining their attack performance in real-world scenarios. We iteratively perform gradient updates on the adversarial patch $(P)$ in the pixel space that optimizes our objective function, as defined below:
\begin{equation}
    L_{total} = \alpha L_{det} + \beta L_{sim} + \gamma L_{tv}
\end{equation}
$L_{det}$ is the adversarial detection loss. $L_{sim}$ is the similarity loss (See Section \ref{stealth}). $L_{tv}$ is the total variation loss on the generated image to encourage smoothness (See Section \ref{robust}).
$\alpha$, $\beta$, and $\gamma$ are hyper-parameters used to scale the three losses. For our experiments we set $\alpha = 1$, $\beta = 4$, and $\gamma = 0.5$.
We optimize the total loss using Adam \cite{adam} optimizer. We try to minimize the object function $L_{total}$ and optimize the adversarial patch. We freeze all weights and biases in the object detector, and only update the pixel values of the adversarial patch. The patch is randomly initialized. 
%============================
%\subsubsection{Adversarial Detection Loss}
%============================
Object detectors, such as YOLO, output an arbitrary number of boxes or detections. For each detection $j$, the goal is to attack two quantities: its objectness probability $D^j_{obj}$ and its class probability $D^j_{cls}$. Minimizing the objectness probability $D^j_{obj}$ causes the j-th object not to get detected. Minimizing class probability $D^j_{cls}$ causes the j-th object to get classified into the wrong class (e.g. person gets classified as a dog.). In this paper, we focus on targeting the person class, e.g., considering the ML-based smart surveillance use cases. Thus, we minimized both the objectness $D^j_{obj}$ and class probabilities $D^j_{cls}$ pertaining to the person class. 
%For faster iterations, we do not compute the loss over all detected boxes. Instead, we only use the detected box having the highest objectness and class probabilities. 
Our adversarial detection loss is defined by: 
\begin{equation}
    L_{det} = \frac{1}{n} \sum^{n}_{i = 1} (\frac{1}{\#objects}\sum^{\#objects}_{j=1} [D^j_{obj} (I'_i)D^j_{cls}(I'_i)])
\end{equation}
%To deflect (deviate) the attention of the model to the naturally looking patch \textcolor{green}{ask victor to update this section}
%============================
\subsection{Attack Stealthiness}
%============================
\label{stealth}

The key idea is to generate a patch similar to an existing image, and this is accomplished by defining a novel loss term that represents the distance of the patch to the original target image. The proposed similarity loss is designed to %minimize the difference (L-norm) between a benign image and the patch.
maximize the cosine similarity between the target benign image and the adversarial patch $P$.
The similarity loss is defined as:  %$L_{sim} = (\frac{1}{n} \sum_{i,j} (\left | P_{i,j} - N_{i,j} \right |) )^2$.
%\textbf{Cosine similarity-based similarity loss}
\begin{equation}
    L_{sim} = -\left(\frac{\sum_{i,j}P_{i,j} N_{i,j}}{\sqrt{\sum_{i,j}P^2_{i,j}}\sqrt{\sum_{i,j}N^2_{i,j}}} \right)^2
\end{equation}

%We square the similarity loss term, to slow the rate of increase (the slope or the rate of change) and delay the convergence of the similarity metric with respect to the detection loss. 
%During the optimization, we have noticed a discrepancies in different loss terms convergence rates: The similarity loss converges significantly faster compared to the detection loss resulting in the total loss getting stuck. So We introduce a non-linearity that slows down the rate of decrease for the similarity loss. We square the difference between the benign image and the adversarial patch in order to mitigate its dominance during optimization, as quadratic functions exhibit slower rates of increase. This approach essentially downweights the impact of the squared term compared to a linear term. 
%This modification can help balance the impact of the terms and prevent one from dominating the optimization process.
During the optimization process, we observed discrepancies in the convergence rates of various loss terms. Specifically, the similarity loss converges notably faster than the detection loss, resulting in a stagnation of the total loss. To address this imbalance, we introduce a non-linear adjustment aimed at slowing down the rate of decrease for the similarity loss. Our approach involves squaring the difference between the benign image and the adversarial patch. This modification serves to mitigate the dominance of the similarity loss during optimization, leveraging the characteristic slower rates of increase exhibited by quadratic functions compared to linear functions. Effectively, this strategy downweights the impact of the squared term, striking a balance between the two loss terms and preventing one from unduly dominating the optimization process. %By incorporating this quadratic term, we aim to refine the optimization dynamics and ensure a more equitable contribution from both the similarity and detection loss terms. This iterative modification enhances the overall stability and convergence of the optimization process.
%By incorporating the quadratic term, we introduce a non-linearity that slows down the rate of increase for the similarity loss. This adjustment is effective in controlling the influence of a particular loss term in relation to others, especially when there are discrepancies in their convergence rates.
%We square the difference between the benign image and the adversarial patch, as in quadratic functions the rate of increase (the slope or the rate of change) is slower which delays the convergence of the similarity metric with respect to the detection loss. In fact, the similarity metric is a linear function compared to the detection metric which is a nonlinear function that converges much slower. 
This similarity loss provides a higher flexibility compared to the GAN-based technique and the limited latent space.
%\begin{equation}
%    L_{sim} = (\frac{1}{n} \sum_{i,j} (\left | P_{i,j} - N_{i,j} \right |) )^2
%\end{equation}
%============================
\subsection{Attack Robustness}
%============================
\label{robust}
Introducing the digital attack into the physical world poses an additional challenge, as the perturbation must be strong enough to withstand real-world distortions arising from variations in viewing distances and angles, lighting conditions, camera limitations, and dynamic objects. Previous studies have revealed that adversarial examples generated via conventional techniques frequently cease to be adversarial after undergoing slight transformations \cite{Luo2015FoveationbasedMA, no_need}. %Attack robustness stands for maintaining attack ability in a dynamic environment, including robust to cross-scenes, robust to physical constraints, etc.
In order to ensure patch robustness we use two preprocessing blocks (i.e., EOT and CT blocks) to generate a large variation of transformed patches to be used in the patch training process. 

%============================
\subsubsection{Expectation Over Transformation (EOT)}
%============================
EOT is a general framework for improving the adversarial robustness of physical attack on a given transformation distribution \textit{T} \cite{eot}. Essentially, EOT is a data augmentation technique for adversarial attacks, which takes potential transformation in the real world into account during the optimization, resulting in better robustness. 
EOT is used to add random distortions in the optimization to make the perturbation more robust. The transformation distribution is presented in Table \ref{transformations}. %\textit{\textbf{Spatial transformations:}} The patch was randomly scaled to a size that is nearly proportional to its actual size in the scene as part of the geometric changes. Rotate the patch $P$ at random ($\pm20^\circ$) about the center of the bounding boxes. The aforementioned replicates printing size and location inaccuracies. %\textit{\textbf{Appearance Transformations:}} The pixel intensity values are altered by adding random noise ($\pm 0.1$), conducting random contrast adjustment of the value ($[\ 0.8, 1.2 ]\ $), and doing random brightness adjustment ($\pm 0.1$) to create the color space conversions. 

%\begin{equation}
%    T_\theta (I_i)
%\end{equation}
%Modeling Deformation of A Moving Object (MDMO)
%\textit{\textbf{Transformation for cloth deformation.}}
\begin{table}[ht]
  \centering
\small
  \begin{tabular}{lcc}
    \hline
    \textbf{Transformations} &  \textbf{Parameters} & \textbf{Remark} \\
    \hline
    Rotation    & $\pm20^\circ$ & Camera Simulation   \\
    %Cropping    & −0.7 $\sim$ 1.0 & Occlude Simulation   \\
    Affine & $0.7$ & Perspective \\
    Scale    & $[0.25, 1.25]$ & Distance/Resize \\
    Random Noise & $\pm0.1$ & Noise   \\
    Brightness  & $\pm0.1$ & Illumination   \\
    Contrast    & $[0.8, 1.2]$ & Camera Parameters   \\
  \hline
\end{tabular}
\caption{Transformation distribution.}
\label{transformations}
\end{table}
%============================
\subsubsection{Creases Transformation (CT) }
\begin{figure}[!htp]
\centering %creases_vis.png
\includegraphics[width=0.33\textwidth]{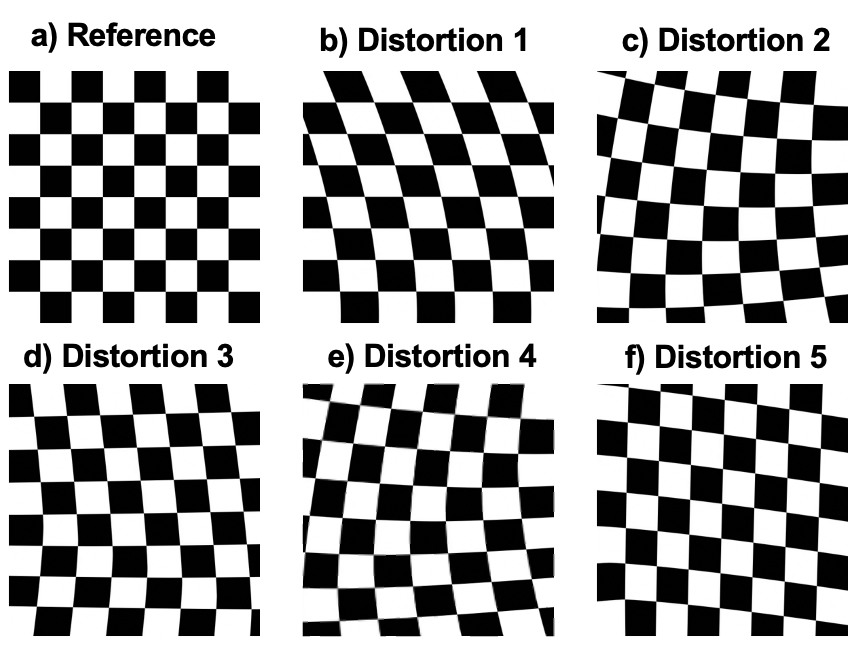} %, height=7cm
\caption{Visual examples of random crease distortion. }
\label{creases_vis}
\end{figure}

%============================
\label{Sec:CT}
Another possible transformation when printing the generated patch on a T-shirt is constantly changing creases in the clothes resulting from a person's movements. To overcome this challenge we propose to perform the following transformations:
Each crease is modeled by randomly selecting a point on the patch, along with a 2D vector representing the crease's direction, which is chosen within a range of 5 degrees to simulate the alignment on the clothing surface. The selected pixels within the patch are displaced in the direction defined by the vector, with varying degrees of movement based on their proximity to the vector's line. This displacement emulates the natural variation in the crease intensity along their length. The displacement of a point $(x, y)$ resulting from the chosen crease point $(x_0, y_0)$ and the associated 2D vector is calculated as Displacement = Vector x Multiplier, where the multiplier captures the dynamic nature of creases. This process is performed for each incorporated crease. 
%A set of creases were added to the patch: Each crease was modeled by randomly selecting a point and a 2D vector ($\pm5$) and moving all pixels in the direction of the vector, with pixels closer to the line formed by the vector extending from the chosen point moving at a greater magnitude. 
The movement of a point $(x, y)$ when the chosen point is $(x_0, y_0)$ 
is the vector multiplied by a multiplier, which follows the equation: %$multiplier(x, y) = 1 - \frac{\sin^2{\theta}[(x-x_0)^2 + (y-y_0)^2]}{width^2 + height^2}$.
\begin{equation}
multiplier(x, y) = 1 - \frac{\sin^2{\theta}[(x-x_0)^2 + (y-y_0)^2]}{width^2 + height^2}
\end{equation}
Where $\theta$ is the angle between the direction of the $(x, y)$ from $(x_0, y_0)$ and the chosen vector, and \textit{width} and \textit{height} are the dimensions of the patch. The resulting transformations are illustrated in figure \ref{creases_vis}.
%The resulting image is forward propagated through the detector. 

%============================
\subsubsection{Total Variation Norm (TV loss)}
%============================
The characteristics of natural images include smooth and consistent patches with gradual color changes within each patch \cite{mahendran2015understanding}. Therefore, 
to increase the plausibility of physical attacks, smooth and consistent perturbations are preferred. Additionally, extreme differences between adjacent pixels in the perturbation may not be accurately captured by cameras due to sampling noise. This means that non-smooth perturbations may not be physically realizable \cite{Sharif2016FaceRecognitionAttacks}. To address these issues, the total variation (TV) \cite{mahendran2015understanding} loss is introduced to maintain the smoothness of the perturbation.
For a perturbation P, TV loss is defined as:
%$L_{tv}$ is the total variation loss on the generated image to encourage smoothness.
%It is defined as:
\begin{equation}
    L_{tv} = \sum_{i,j} \sqrt{(P_{i+1,j} - P_{i,j})^2 + (P_{i,j+1} - P_{i,j})^2}
\end{equation}
%$L_{tv} = \sum_{i,j} \sqrt{(P_{i+1,j} - P_{i,j})^2 + (P_{i,j+1} - P_{i,j})^2}$.
Where the subindices $i$ and $j$ refer to the pixel coordinate of the patch $P$.
%%%%%%%%%%%%%%%%%%%%%%%%%%%%%%%%%%%%%%%%%%%%%%%%%%%%%%%%%%%%%%%%%%%%%%%%%%%%%%%%

%============================
\section{Evaluation of Attack Performance}
%============================
\subsection{Experimental Setup}
\label{sec:setup}
As victim object detectors, we used Yolov2 \cite{yolov2}, Yolov3 \cite{yolov3}, Yolov3tiny, Yolov4 \cite{yolov4}, Yolov4tiny, Yolov7 \cite{yolov7} and FasterRCNN \cite{faster} with an input image resolution of $416 \times 416$. For the optimization, we use Adam optimizer with a learning rate of $0.001$, $\beta1=0.9$ and $\beta2=0.999$.
In our evaluations, we use the images of the INRIA \cite{inria} person dataset with pedestrians, which makes it suitable for our proposed use case setting. The dataset is popular in the Pedestrian Detection community, both for training detectors and reporting results. It consists of 614 person detections for training and 288 for testing.
%We conducted our experiments on a Linux-based PC with an NVIDIA GeForce GTX 3060 Ti. The models were implemented using Python version 3.10.12 and PyTorch \cite{NEURIPS2019_9015} version 2.1.0.

\label{sec:eval}
%============================
\subsection{DAP Transferability}
%============================
Our evaluation metric is the mean average precision (mAP), which is a commonly used performance measure in object detection tasks. To calculate mAP, we adopt the approach used in prior work \cite{thys2019, Hu21}: we take the detection boxes generated by each detector on the clean dataset as the ground truth boxes (assuming no adversarial patch is present), and then report the mAP when the same detectors are used to detect objects with the added adversarial patches.
Table \ref{performance_} presents the evaluation results on the INRIA dataset, where we trained our adversarial patches using four different detectors. The victim detectors used during testing are shown in the horizontal rows, while the detectors used during training are shown in the vertical columns. As shown in the table, our approach achieves lower mAP scores across various detector combinations, demonstrating its effectiveness and transferability.
\begin{table}[!htp]
  \centering
\small
  \begin{tabular}{lcccc}
    \toprule
    \textbf{Detector} & \textbf{Yolov3}  & \textbf{Yolov3tiny} & \textbf{Yolov4}  & \textbf{Yolov4tiny} \\
    \midrule
      \textbf{Yolov3}       &  \textbf{32.63$\%$} & 37.13$\%$  &  44.31$\%$  &  38.08$\%$ \\
      \textbf{Yolov3tiny}   &  35.93$\%$ & \textbf{6.54$\%$ } &  43.96$\%$ &  35.21$\%$ \\
      \textbf{Yolov4}       & 50.21$\%$  &  51.33$\%$ & \textbf{24.65$\%$}  &  48.8$\%$\\
      \textbf{Yolov4tiny}   & 41.36$\%$  & 26.47$\%$  & 45.18$\%$ & \textbf{16.98$\%$}\\
  \bottomrule
\end{tabular}
   \caption{\label{performance_} Attack performance in terms of mAP of DAP on INRIA dataset using different detectors.}
\end{table}

Our experimental results, presented in Table \ref{performance_}, demonstrate that when using the same victim detector for training and testing, our attack achieves high performance ($93.46\%$ attack success rate when attacking Yolov3tiny). Furthermore, our technique exhibits strong transferability across different victim detectors ($64.07\%$ attack success rate when attacking Yolov3 using a patch trained using Yolov3tiny), as evidenced by its success against models that were not used during training. This finding highlights the transferability of our approach.
%=============================
\subsection{DAP vs. State-of-the-Art Techniques}
%=============================
\label{sec:comp}
To assess the performance of our proposed adversarial patch, we compare it against four state-of-the-art techniques: Naturalistic Patches \cite{Hu21}, Adversarial T-shirt \cite{adversarialtshirt}, Adversarial YOLO \cite{thys2019}, and Universal Physical Camouflage (UPC) \cite{Huang2020}. Table \ref{comparison} summarizes the mean average precision results of these methods on the INRIA dataset. Our proposed approach achieves competitive attack performance compared to the state-of-the-art techniques.
Notably, Adversarial YOLO \cite{thys2019} delivers the highest attack performance for most of the detectors, but the generated patches lack realism and can be easily detected due to their noticeable appearance. In contrast, our approach generates adversarial patches that blend more naturally into the surrounding while maintaining high attack performance. %Our proposed patch features realistic patterns that can help to conceal their adversarial nature. 
Overall, our approach delivers comparable performance to that of state-of-the-art techniques, while offering a more natural and subtle visual appearance.
%\begin{wrapfigure}{l}{10.5cm}
\begin{table}[!ht]
  \centering
\small
  \begin{tabular}{lcccc}
    \toprule
    %\textbf{} & \textbf{}& \textbf{} & \textbf{ } & \textbf{}\\
    \textbf{Detector} & \textbf{DAP }& \textbf{NAP}  &\textbf{Adv. YOLO }& \textbf{UPC}\\
    \midrule
      \textbf{Yolov2}      &  19.51$\%$    &   12.06$\%$   & 2.13$\%$ & 48.62$\%$\\
      \textbf{Yolov3}      &  32.63$\%$  &   34.93$\%$  & 22.51$\%$  &   54.40$\%$\\
      \textbf{Yolov3tiny}  &  6.54$\%$ & 10.02$\%$  &  8.74$\%$&     63.82$\%$\\
      \textbf{Yolov4}      &  24.65$\%$ &   22.63$\%$ & 12.89$\%$&   64.21$\%$\\
      \textbf{Yolov4tiny}  &  16.98$\%$ &   8.67$\%$ &3.25$\%$&   57.93$\%$\\
      \textbf{Yolov7}  &  17.72$\%$ &   60.78\% & N/A &   N/A \\
  \bottomrule
\end{tabular}
  \caption{\label{comparison}DAP vs State-of-the-Art adversarial patches \textbf{without transformations}.}
\end{table}
%\end{wrapfigure}
%=============================
\subsection{Impact of Adding Non-Rigid Deformations on Patch Performance}
%=============================
\label{wrinkles}

To evaluate the robustness of different adversarial patch techniques to variations in clothing appearance, we applied random clothes deformation transformations to the INRIA dataset and measured the attack success rates. We compared the performance of our proposed DAP patch against that of the state-of-the-art Naturalistic Patches \cite{Hu21}. Table \ref{performance_transformations} summarizes the results.
We found that the effectiveness of the NAP was drastically degraded after applying the clothes deformation transformations. For example, the success rate of the Yolov3tiny-based NAP dropped from $89.98\%$ to $30.8\%$. In contrast, our DAP patch maintained its effectiveness against the deformed clothing, with only a minor drop of $8.9\%$ in the success rate.
These results suggest that our proposed DAP patch is more robust and less affected by variations in clothing appearance modeled by the creases in the patch compared to the state-of-the-art Naturalistic Patches. %This robustness can be particularly important in real-world scenarios where clothing appearance can vary widely, making it challenging to generate effective adversarial patches.
Our proposed patch continues to demonstrate remarkable performance on the FasterRCNN model, surpassing the capabilities of the state-of-the-art NAP method. In fact, in the presence of non-rigid transformations, our DAP approach maintains a mean average precision (mAP) of 30.60\%, outperforming NAP, which experiences a decrease in performance through an increase in the mAP from 42.47\% to 75.1\%.

\begin{table}[htp]
\centering
\small
    \begin{tabular}{lcccc}
    \toprule
    \textbf{} &  \multicolumn{2}{c}{\textbf{NAP}} & \multicolumn{2}{c}{\textbf{DAP}}\\
    \cmidrule(r){2-3} \cmidrule(r){4-5}
    \textbf{Detector} & \textbf{w/o}  & \textbf{w/}  &  \textbf{w/o}   & \textbf{w/} \\
    \midrule
     \textbf{Yolov3}       &  $34.93\%$ &  $77.37\%$  & \textbf{32.63$\%$ }  & \textbf{37.70$\%$}\\
     \textbf{Yolov3tiny}   & $10.02\%$  &  $69.20\%$  & \textbf{6.54$\%$}   &  \textbf{15.44$\%$}\\
     \textbf{FasterRCNN}   & $42.47\%$  &  $75.1\%$  & \textbf{19.19$\%$ }  & \textbf{30.60$\%$} \\
     \textbf{Yolov7}   & 60.78\%  & 72.46\%   & \textbf{17.72$\%$ }  &\textbf{ 36.67$\%$} \\
     \bottomrule
  \end{tabular}
  \captionof{table}{\label{performance_transformations}Attack performance (mAP) of different Patches \textbf{with and without rigid and non-rigid transformations}.}
\end{table}

%============================
\section{Physical Attack Evaluations}
%============================

%============================
%\subsection{DAP-based T-shirt}
%============================
For physical attack evaluation and to compare the effectiveness of our adversarial patch (DAP) and the naturalistic patch (NAP), both generated using Yolov3tiny, we printed them on T-shirts in 20.5cm x 21.5cm format. %As shown in Figure \ref{DAP_vs_commercial}, our generated patches exhibit a natural and striking resemblance to commercially available T-shirts. The visual similarity between our patches and genuine T-shirt designs is evident, highlighting the seamless integration of our adversarial patches into everyday attire. This visual appeal and resemblance to commercial products are crucial for the practical implementation of our approach, as it ensures inconspicuousness and reduces the likelihood of detection by a human observer. %These findings further support the effectiveness and potential real-world utility of our Dog-based DAP and reinforce the significance of our research in the domain of adversarial patch generation.
%For physical attack evaluation, we carried out our real-world experiments using a set of 100 test samples, each featuring a person wearing and adversarial t-shirt: We print the generated patches in 20.5cm x 21.5cm format on a T-shirt. We use Yolov3tiny to generate adversarial patches for experiments.
%We quantified attack performance based on attack success rate.
%Figure \ref{} illustrates the impact of our adversarial T-shirts on detection recall compared to benign ones. Our two patches succeeded in hiding the person in spite of the creases on the T-shirts. % but on the other hand, the patch of \cite{Hu21} failed to deceive the detector.
%Looks natural, similar to commercialized T-shirts..
%\label{physical}
%\begin{wrapfigure}{l}{5cm}
%\includegraphics[width=5cm, height=3cm]{figures/commercial.pdf}
%\caption{DAP-based T-shirts vs Commercialized T-shirts. Our DAP-based T-shirt seems indistinguishable from other commercialized ones}\label{DAP_vs_commercial}
%\end{wrapfigure}
We filmed two videos to showcase the effectiveness of our DAP-based T-shirt and that of the GAN-based T-shirt in fooling the Yolov3tiny object detector (The two demo videos are provided in the supplementary material). The participant is approximately three meters away from the camera. We asked the participant to move in different directions: back and forth as well as side-to-side within the duration of the videos. The participant hid the patch using his hand forming the baseline performance of the detector and also deliberately created aggressive wrinkles to the patches to assess their robustness to different transformations.

The detection results were recorded in these videos. Subsequently, we extracted frames from the videos and performed annotations using the following procedure:
\begin{itemize}
    \item Patch Presence (P) Annotation: If a patch is detected in a frame, we assign a value of $P=1$. If no patch is present, we set $P=0$.
    \item Person Detection (D) Annotation: We determine whether a person is detected in the frame or not. If a person is detected, we assign $D=1$. If no person is detected, we set $D=0$.
\end{itemize}

\begin{figure}[ht]
\centering
\includegraphics[width=0.9\columnwidth]{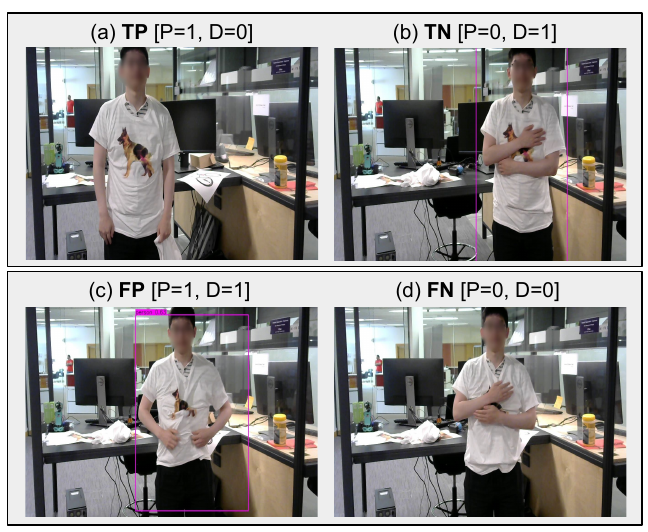}
\caption{Example of annotated samples.}
\label{annotation}
\end{figure}

To illustrate this annotation process, Figure \ref{annotation} showcases four different cases. The key metrics used to evaluate the performance of the detection system are as follows:
The True Positive Rate (TPR), the False Positive Rate (FPR), the True Negative Rate (TNR), and the False Negative Rate (FNR) (Further details about these metrics and the way they were computed is provided in the supplementary material). These evaluation metrics allow us to assess the performance of our adversarial patches and provide quantitative insights into the detection results obtained during our experiments.

%As presented in Table \ref{rates}, our patch, DAP, demonstrates the highest \textbf{true positive rate}, indicating that in 92.01\% of the cases where our patch was present and the person was not detected. In comparison, the NAP patch achieved a success rate of only 57.22\%. In addition, our patch exhibited the lowest \textbf{false positive rate}, with only 8.33\% of cases falsely detecting the presence of a person when our patch was actually present. In contrast, the NAP patch had a significantly higher false positive rate of 42.77\%, incorrectly identifying the presence of a person despite the patch being present.
 \begin{table}[!ht]
  \centering
\small
  \begin{tabular}{ccccc}
   \toprule
    \textbf{Patch} &\textbf{TPR}&  \textbf{TNR}&  \textbf{FPR}&  \textbf{FNR}\\

    \midrule
      \textbf{DAP}    & \textbf{92.01$\%$}   & \textbf{ 59.66$\%$}   &  \textbf{8.33$\%$}   &   \textbf{40.33$\%$ }  \\
      \textbf{NAP}  & 57.22$\%$  &  43.32$\%$   &  42.77$\%$  &   56.67$\%$ \\
  \bottomrule
\end{tabular}
  \caption{\label{rates} TPR, TNR, FPR, and FNR for DAP and NAP-based adversarial T-shirts}  %Confusion matrix
\end{table}
%& \textbf{PPV}
%\textcolor{red}{To Check}
Table \ref{rates} provides a comprehensive overview of the evaluation metrics for our adversarial patch, DAP, as well as the naturalistic Patch (NAP). These metrics highlight the performance of the patches in terms of true positive rate (TPR) and false positive rate (FPR). For instance, our DAP patch demonstrates an impressive TPR of 92.01\%, indicating that in 92.01\% of cases where the person was not detected was because of our patch and 7.99\% because of the detector not working well. On the other hand, the NAP patch achieved a considerably lower success rate of only 57.22\% of the cases the miss-detection of the person was because of the patch. 
%Furthermore, our DAP patch exhibited the lowest FPR, with only 8.33\% of cases falsely detecting the presence of a person when our patch was actually present. In contrast, the NAP patch had a significantly higher FPR of 42.77\%, incorrectly identifying the presence of a person despite the patch being present. These findings clearly demonstrate the superior performance of our DAP patch in terms of concealing the presence of the targeted object (person) and minimizing false positive identifications, highlighting its effectiveness in circumventing detection systems compared to the NAP patch.
%When evaluating the detector's performance in a benign scenario without adversarial patches, we observed notable differences between the DAP and NAP settings. Notably, when the patch was covered, the detecor exhibited superior performance in detecting the presence of a person. Specifically, we observed a higher true negative rate (TNR) of 59.66\% for the DAP setting. In comparison, in the NAP setting, the detector achieved a lower TNR of 43.33\%. Furthermore, the DAP setting demonstrated a lower false negative rate (FNR), indicating that the detector performed better in detecting persons in the benign scenario. This finding highlights that our patch was evaluated in stronger settings. 
%Additionally, our patch achieves the lowest FPR corresponding to lower percentage of the person detected (i.e., 8.33\%) while the patch is visible compared to a percentage of 42.77\% of the times where the patch was visible and it didn't work (i.e., the person was detected).
Furthermore, our patch achieves the lowest false positive rate (FPR) of 8.33\%, where the person was detected while the patch was visible. This contrasts with the 42.77\% occurrence where the patch was visible, but it failed to escape detection.
%In summary, our DAP patch consistently outperformed the NAP patch in terms of true positive rate (TPR) and false positive rate (FPR) even in a scenario where the detector's performance was generally better. This emphasizes the superior results achieved by our proposed patch in successfully deceiving the detector and underscores its effectiveness in evading detection systems.
Further details and discussion can be found in the supplementary material.
\section{Discussions}
%------------------------------------------
\subsection{Ablation Study}
%------------------------------------------
%Similarity Loss Ablation: Without the similarity loss function, the generated patch effectively becomes noise with conspicuous patterns. This highlights the pivotal role of the similarity loss in guiding the patch to resemble a benign image and enhancing its inconspicuousness. Detection Loss Ablation: Without the detection loss, we find that the generated patch becomes remarkably similar to the target image. However, interestingly, this change does not lead to any discernible impact on the detector's performance. This suggests that the detection loss plays a crucial role in ensuring the patch's adversarial nature and its effectiveness in evading detection. Total Variation Loss Ablation: Without the total variation loss function, we observe that the color change within the patch becomes sharper. This indicates that the total variation loss contributes significantly to the smoothness and natural appearance of the generated patch. We will make sure to add this study to the revised manuscript.
We conducted an ablation study for each term in our loss function.\\
\noindent \textit{Similarity Loss Ablation}: When the similarity loss function is removed, the generated patch essentially turns into noisy, conspicuous patterns but more efficient patch achieving a mAP of $2.54\%$ for Yolov3tiny. This underscores the vital role of the similarity loss in guiding the patch to resemble a benign image and enhancing its inconspicuousness.\\
\noindent \textit{Detection Loss Ablation}: In the absence of the detection loss, the generated patch closely resembles the target image. As expected, this change doesn't seem to noticeably impact the detector's performance. This suggests that the detection loss plays a critical role in maintaining the patch's adversarial nature and its ability to evade detection.\\
\noindent \textit{Total Variation Loss Ablation}: Removing the total variation loss function results in sharper color changes within the patch. This indicates that the total variation loss significantly contributes to the smoothness and natural appearance of the generated patch.

%------------------------------------------
\subsection{Impact of Patch Scale}
%------------------------------------------
%To evaluate the influence of patch size on the effectiveness of our proposed adversarial patch, we conducted a series of digital experiments on the INRIA dataset. Specifically, 
We evaluated the patch performance with respect to the size of the target object (person). We use scales of $0.6, 0.5, 0.4,$ and $0.3$ representing the size of the patch with respect to the size of the person bounding box ($0.5$ scale corresponds to $0.2$ in \cite{Hu21}). Our experimental results illustrated in Table \ref{patch_size}, confirm that larger patches generally result in stronger attack performance, as expected. This is due to the fact that a larger patch covers more of the target object, making it more difficult for object detectors to identify the person. Overall, our experimental findings suggest that patch size is an important factor to consider when designing effective adversarial patches for object detection.

\begin{table}[htp]
\small
\centering
    \begin{tabular}{ccccc}
    \toprule
    \textbf{Scale} & \textbf{0.3} & \textbf{0.4}& \textbf{0.5} & \textbf{0.6} \\
    \midrule
      \textbf{mAP}  &   58.98$\%$  &   37.20$\%$ &   15.44$\%$ &   6.54$\%$   \\
     \bottomrule
  \end{tabular}
  \caption{\label{patch_size}Attack performance against YOLOv3tiny with adversarial patches in different scales with respect to the target object size for the INRIA dataset.}
\end{table}

%\begin{table}[!ht]
%\begin{wrapfigure}{r}{6cm}
%\small
%\centering
%    \begin{tabular}{cc}
%    \toprule
%    \textbf{Patch size} & \textbf{mAP}\\
%    \midrule
%      \textbf{0.3}  &   58.98$\%$     \\
%      \textbf{0.4}  &   37.20$\%$     \\
%      \textbf{0.5}  &   15.44$\%$   \\
%      \textbf{0.6}  &   6.54$\%$   \\
%     \bottomrule
%  \end{tabular}
%  \caption{\label{patch_size}Attack performance against YOLOv3tiny with adversarial patches in different scales with respect to the target object size for the INRIA dataset.}
%\end{wrapfigure}

%\begin{wrapfigure}{l}{5cm}
%\centering
%\small
%    \begin{tabular}{cc}
%    \toprule
%    \textbf{Patch size} & \textbf{mAP}\\
%    \midrule
%      \textbf{0.3}  &   58.98$\%$     \\
%      \textbf{0.4}  &   37.20$\%$     \\
%      \textbf{0.5}  &   15.44$\%$   \\
%      \textbf{0.6}  &   6.54$\%$   \\
%     \bottomrule
%  \end{tabular}
%  \captionof{table}{ \label{patch_size}Attack performance against YOLOv3tiny with adversarial patches in different scales with respect to the target object size for the INRIA dataset}
%\end{wrapfigure}
%\begin{wrapfigure}{l}{6cm}
%\includegraphics[width=6cm,height=5cm]{figures/size.pdf}
%\captionof{figure}{\label{Fig:size}Detection results for different patch scales.}
%\end{wrapfigure}
%------------------------------------------
\subsection{Adversarial Patches in Different Classes}
%------------------------------------------
Our proposed adversarial patch is not limited to targeting only the "Dog" class. In fact, we were able to successfully generate effective patches for other object classes as well, such as the "Cat" and "Bike" classes. This demonstrates the versatility and potential of our approach, which can be applied to a wide range of target patterns of patch appearance. %With further development and optimization, our proposed method could be used to generate effective adversarial patches for a variety of object classes and scenarios.
\begin{figure}[!ht]
\centering
\includegraphics[width=\columnwidth]{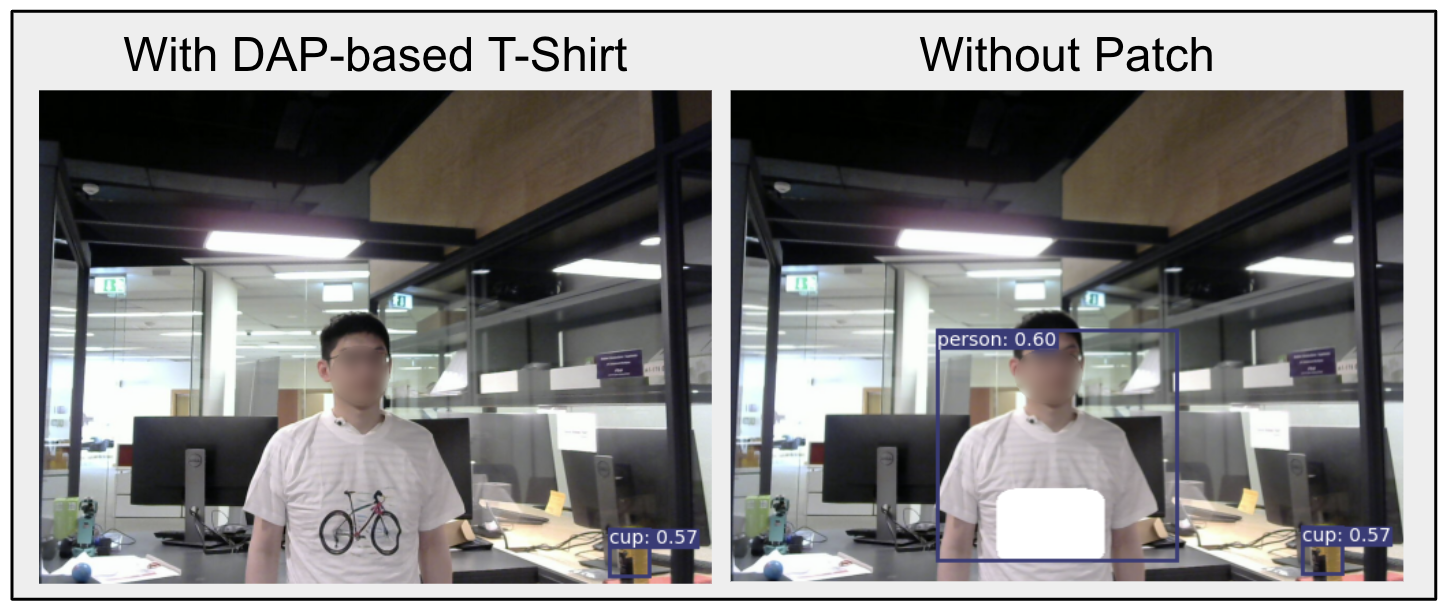}
\caption{Detection results for other classes.}
\label{result_10}
\end{figure}
%\label{physical}
Figure \ref{result_10} shows the detection results when printing the generated patches and the patches performance is summarized in Table \ref{classes}. For instance, when using a Bike as the target pattern, and when applying creases to the generated patch we get 75.95\% ASR. % we achieve 75.95\% attack success rate.
\begin{table}[ht]
\centering
\small
    \begin{tabular}{lcccc}
    \toprule
    \textbf{} & \multicolumn{2}{c}{\textbf{Bike Class}} & \multicolumn{2}{c}{\textbf{Cat Class}}\\
    \cmidrule(r){2-3} \cmidrule(r){4-5}
      \textbf{Detector} & \textbf{w/o} & \textbf{w/} & \textbf{w/o} & \textbf{w/}\\
    \midrule
      \textbf{Yolov3}      &   45.46$\%$  &  47.58$\%$&  38.87$\%$ & 47.25$\%$\\
      \textbf{Yolov3tiny}  &   18.43$\%$  &  24.05$\%$&   42.67$\%$&  50.05$\%$ \\
     \bottomrule
  \end{tabular}
  \captionof{table}{ \label{classes} DAP performance with and without transformations for the 'Bike' and 'Cat' classes for Yolov3 and Yolov3tiny detectors.}
\end{table}

%------------------------------------------
%\subsection{Ethical Considerations}
%------------------------------------------
%A fundamental goal of our work is to raise awareness among users about potential vulnerabilities inherent in existing algorithms, which can be exploited by adversaries to achieve ulterior motives. This work will help the users (society in general) understand the limitations of the existing (deep learning-based) solutions and encourage them to make educated decisions instead of blindly trusting the technology for safety and security-critical applications such as surveillance. Apart from that, our work will encourage the deep learning and system design community to focus on building robust systems/models for real-world use cases.
%------------------------------------------
\subsection{Limitations}
%------------------------------------------
Enhancing the interpretability and explainability of our approach is crucial for gaining deeper insights into the underlying mechanisms that contribute to its success. By understanding the specific features or characteristics that make certain target classes result in more effective patches, we can refine our approach to improve its performance across a wider range of classes. Exploring techniques such as feature visualization, attribution methods, or saliency analysis can help us identify the discriminative patterns that our approach leverages to deceive the detector. This knowledge can guide the development of more effective and generalizable adversarial patches.
\section{ Related Work}
\label{sec:related}
Initially, physical attacks aimed at fooling person detectors were generated without considering patch stealthiness, but rather focused on performance and producing effective attacks. For example, \cite{thys2019} proposed printable adversarial patches attached to a cardboard, \cite{Lee2019} proposed patches trained for random placement on the scene, and authors in \cite{Wu2020} proposed an invisibility cloak. However, these patches had conspicuous and easily identifiable patterns. To overcome this limitation, some works proposed leveraging the learned image manifold of pre-trained GANs upon real-world images to create naturalistic patches  \cite{Hu21, doan2022tnt, Pavlitskaya22}. Authors in \cite{Huang2020} also proposed a universal camouflage pattern that is visually similar to natural images and for stealthiness added a $L_\infty$ norm constraint to control the adversarial noise. These attacks were aimed to be printed on a T-shirt, but they ignored non-rigid deformations caused by a moving person. Authors in \cite{adversarialtshirt} attempt to model these deformations using a thin plate spline (TPS) based transformer. 
Nevertheless, this method ignores the stealthiness of the patch and results in conspicious patterns. To address all these limitations and solve the trilemma of efficiency, stealthiness, and robustness, we propose an attack that generates a patch, DAP, that maintains naturalistic patterns, is robust to multiple transformations, and can be printed on a T-shirt while being stealthy. A comparison of DAP with state-of-the-art attacks is provided in Table \ref{Tab:related}.
\begin{table}[!ht]
  \centering
\small
  \begin{tabular}{lcccc}
    \toprule
    \textbf{}              & \textbf{Robustness } & \textbf{Stealthiness}& \textbf{} &\textbf{}\\
    \textbf{Attack}       & \textbf{Techniques} & \textbf{Techniques}& \textbf{Object} & \textbf{Space}\\
    \midrule
      \textbf{\cite{thys2019} }  & EOT, TV, NPS  &  N/A & Static & 2D \\
      \textbf{\cite{adversarialtshirt}} & EOT, TPS    & N/A  & Dynamic & 2D\\ 
     \textbf{\cite{Huang2020} }            &  EOT, TV     &  $L_\infty$ norm & Static & 3D \\
      \textbf{\cite{Hu21}}       & TV          & GAN  & Static& 2D\\
      \textbf{Ours}         & EOT, TV, CT  & $L_{sim}$  & Dynamic& 2D \\
  \bottomrule
\end{tabular}
\caption{\label{Tab:related}DAP vs State-of-the-Art adversarial patches in terms of robustness techniques, stealthiness techniques, the targeted objects, and the Space. (EOT: Expectation Over Transformations, TV: Total Variation, NPS: Non-Printability Score, TPS: Thin Plate Spline, CT: Creases Transformation).}
\end{table}

\section{Conclusion}
\label{sec:conclusion}
This paper presents a novel approach to generate naturalistic physical adversarial patches for object detectors. Our proposed framework overcomes the limitations of GAN-based approaches including the limited latent space, producing stealthy patches that achieve competitive attack performance. 
Furthermore, we introduce a creases transformation blocks to model non-rigid deformations aimed for dynamic objects. Our approach results in an effective, stealthy, and robust adversarial patch. %, DAP,  that stands out from existing non-naturalistic techniques.
{
    \small
    \bibliographystyle{ieeenat_fullname}
    \bibliography{main}
}


\end{document}